\documentclass[12pt]{article}
\usepackage{graphicx}
\usepackage{amsmath,verbatim,epsfig}
\usepackage{epstopdf}
\usepackage[utf8x]{inputenc}
\usepackage[colorlinks=true,linkcolor=blue,citecolor=blue]{hyperref}
\usepackage[usenames,dvipsnames]{color}
\usepackage{bm}

\newcommand\pubnumber{
}
\newcommand\pubdate{\today}

\newcommand{\nn}{\nonumber}

\newcommand{\be}{\begin{equation}}
\newcommand{\ee}{\end{equation}}
\newcommand{\bea}{\begin{eqnarray}}
\newcommand{\eea}{\end{eqnarray}}
\newcommand{\balign}{\begin{align}}
\newcommand{\ealign}{\end{align}}
\newcommand{\as}{\alpha_s}

\newcommand{\bg}{\begin{gather}}
\newcommand{\foma}{\end{gather}}

\newcommand{\noopsort}[1]{}

\newcommand{\vecb}[1]{\mbox{\boldmath $#1$}}

\def\z{\zeta}

\def\<{\langle}
\def\>{\rangle}

\def\a{\alpha}
\def\b{\beta}
\def\g{\gamma}  \def\G{\Gamma}
\def\d{\delta}  
\def\l{\lambda}

\def\m{\mu}

\def\z{\zeta}

\def\({\left(}
\def\[{\left[}
\def\){\right)}
\def\]{\right]}

\def\ln{\hbox{ln}}

\def\Dslash{D\!\!\!\!\slash}

\def \le { \left    }
\def \ri { \right }

\def\cagliari{Dipartimento di Fisica, Universit\`a di Cagliari and INFN, Sezione di Cagliari, I-09042, Monserrato (CA), Italy
}
\def\nikhef{Nikhef and Department of Physics and Astronomy,
VU University Amsterdam,\\
De Boelelaan 1081, NL-1081 HV Amsterdam, the Netherlands
}
\def\torino{
Dipartimento di Fisica, Universit\`a di Torino, Via P. Giuria 1, \\
I-10125 Torino, Italy}

\def\madrid{Departamento de F\'isica Te\'orica II,
Universidad Complutense de Madrid (UCM),
28040 Madrid, Spain}

\def\support{\footnote{Speaker.}}

\def\Title#1{\begin{center} {\Large #1 } \end{center}}
\def\Author#1{\begin{center}{ \sc #1} \end{center}}
\def\Address#1{\begin{center}{ \it #1} \end{center}}

\newcommand\pubblock{\rightline{\begin{tabular}{l} \pubnumber\\
         \pubdate  \end{tabular}}}
\newenvironment{Abstract}{\begin{quotation}  }{\end{quotation}}
\newenvironment{Presented}{\begin{quotation} \begin{center}
             PRESENTED AT\end{center}\bigskip
      \begin{center}\begin{large}}{\end{large}\end{center} \end{quotation}}
\def\Acknowledgements{\bigskip  \bigskip \begin{center} \begin{large}
             \bf ACKNOWLEDGEMENTS \end{large}\end{center}}

\def\beq{\begin{equation}}
\def\eeq#1{\label{#1}\end{equation}}
\def\eeqn{\end{equation}}

\def\beqa{\begin{eqnarray}}
\def\eeqa#1{\label{#1}\end{eqnarray}}
\def\eeqan{\end{eqnarray}}

\let\bar=\overbar

\def\Dslash{\not{\hbox{\kern-4pt $D$}}}
\def\dslash{\not{\hbox{\kern-2pt $\del$}}}

\def\ee{e^+e^-}

\def\msb{{\bar{\ssstyle M \kern -1pt S}}}

\begin{document}
\begin{titlepage}
\pubblock

\vfill
\Title{Perturbative and non-perturbative QCD effects in transverse momentum distributions}
\vfill
\Author{ U. D'Alesio$^a$, M.G. Echevarria$^b$, S. Melis$^c$, I. Scimemi$^d$\support}
\Address{$^a$\cagliari \\
$^b$\nikhef\\
$^c$\torino\\
$^d$\madrid}
\vfill
\begin{Abstract}
The recent formulation of the factorization theorem for transverse momentum spectra in terms of well-defined transverse momentum dependent distributions (TMDs), allows for a better understanding of the role of the perturbative and non-perturbative QCD contributions and their interplay. In particular, non-perturbative effects are often included both in the evolution  of the TMDs and in the modeling of their scale-independent part.
Using the available Drell-Yan and $Z$-production data at low and high energy we show that, for the currently available dilepton invariant mass values, the TMD evolution is driven mainly by its perturbative contribution, while the needed non-perturbative correction is scale independent. We then detail the difference between the scale dependences in the high and low boson transverse momentum regime and discuss the related theoretical errors.
\end{Abstract}
\vfill
\begin{Presented}
CIPANP 2015\\
Vail (CO),  May 19--24, 2015
\end{Presented}
\vfill
\end{titlepage}
\def\thefootnote{\fnsymbol{footnote}}
\setcounter{footnote}{0}

\section{Introduction}
Multi-differential cross sections at hadron colliders require a  good control of Quantum Chromo Dynamics (QCD) for a  correct understanding of the experimental results, which are usually run at different center of mass energies, luminosities, etc..
The non-perturbative QCD effects can be included into well-defined objects which are not simply integrated parton distribution functions (PDFs) or fragmentation functions (FFs).
The recently formulated factorization theorems for transverse momentum spectra in Drell-Yan (DY), Semi Inclusive Deep Inelastic Scattering (SIDIS) and $e^+e^-\rightarrow 2$ hadrons \cite{Collins:2011zzd,GarciaEchevarria:2011rb,Echevarria:2012js,Echevarria:2014rua} allow to express the differential cross sections as a convolution of Transverse Momentum Dependent Distributions (TMDs) in momentum space. In the case of DY we  schematically have
\begin{align}
\frac{d\sigma}{dQ dq_T} &\sim
H(Q^2,\mu^2)\\ \times\nn &
\int d^2\vecb k_{AT}\, d^2\vecb k_{BT} \,
F_A(x_A,\vecb k_{AT};\z_A,\m)\,
F_B(x_B,\vecb k_{BT};\z_B,\m)\,
\d^{(2)}(\vecb k_{AT}+\vecb k_{BT}-\vecb q_{T})\,,
\end{align}
where the $F$'s are the TMD parton distribution functions (TMDPDFs), which incorporate all the non-perturbative QCD information, $\zeta_{A, B}\sim {\cal O}(Q^2)$ with $\zeta_A\zeta_B=Q^4$
and $x_{A,B}$ the Bjorken variables. In  coordinate space, or $\vecb b_T$-space, the cross section is proportional to a product of TMDs and $\vecb b_T$ is the variable Fourier conjugate to $\vecb q_{T}$, the dilepton (or virtual boson) transverse momentum.  At the moment the only way to extract the non-perturbative part of TMDs is based on data analysis, and this should be consistent with the perturbative and calculable part of the TMDs.
It is often assumed that the non-perturbative  part of the TMDs is  encoded both  in the evolution kernel, and  in the intrinsic scale-independent part of the TMDs.
In this contribution we comment on our recent analysis of DY and $Z$-boson production data and the information that we can extract~\cite{D'Alesio:2014vja}.
This study makes evident some relevant outcomes. The first result is that for data with dilepton invariant masses above 4 GeV the so-called non-perturbative part of the evolution kernel is not essential for the fit: in other words, one can work solely with a perturbatively-resummed  evolution kernel with no model assumptions. All model dependence is therefore scale independent and can be expressed in terms of 2 parameters. The other relevant aspect is the role of perturbative scales and the error induced  by their variation.
In this respect the data at the $Z$-scale and the low-energy DY data have a different behavior. This supports the idea that a TMD framework is necessary in order to have a comprehensive understanding of data at all energy scales.
In the next section we briefly review the formalism for the easiness of the reader and discuss the origin of scale dependences and theoretical errors, while we refer to the original work~\cite{D'Alesio:2014vja} for a more extensive and detailed presentation. The comparison with available data is done in sections \ref{sec:Z}-\ref{sec:DY}. Conclusions are in section~\ref{sec:conclusions}.

\section{Evolution and Operator Product Expansion of TMDPDF}
\label{sec:evolution}
In order to have a reliable TMD factorization of cross sections, to be compared with data, one needs to implement the TMD evolution kernel  and give a parametrization for TMDs such that their perturbative $q_T$ limit is also recovered (e.g. the part that is usually referred to as the operator product expansion of a TMDPDF onto a PDF).
Here we just collect the relevant formulas for DY and vector boson production in order to focus on the main discussion. A more detailed treatment is given in the original papers~\cite{Collins:2011zzd,GarciaEchevarria:2011rb,Echevarria:2012js,Echevarria:2014rua,D'Alesio:2014vja}.

In coordinate space  the TMDPDF can be written as
\begin{align}
\label{eq:FRF}
\tilde F(x,b_T;\z_f,\m_f)=\tilde R(b_T;\z_i,\m_i,\z_f,\m_f) \tilde F(x,b_T;\z_i,\m_i)
\,,
\end{align}
where $\tilde R$ is the evolution kernel,
\begin{align}\label{eq:tmdkernel}
\tilde R(b_T;\z_i,\m_i,\z_f,\m_f) &=
\exp\le\{
\int_{\m_i}^{\m_f} \frac{d\bar\m}{\bar\m}
\g_F\le(\as(\bar\m),\ln\frac{\z_f}{\bar\m^2} \ri)
\ri\}
\le( \frac{\z_f}{\z_i} \ri)^{-D\le(b_T;\m_i\ri)}
\,.
\end{align}
In this equation $\g_F$ is the anomalous dimension of the TMDPDF and is related to the anomalous dimension of the hard factor as explained in Ref.~\cite{GarciaEchevarria:2011rb}.
The function $D$ obeys  the renormalization group equation
\begin{align}
\label{eq:Deq}
\frac{d D}{d\ln\mu}&=\G_{\rm cusp}\,,
\end{align}
where $\G_{\rm cusp}$ is the cusp anomalous dimension.
The solution of this  differential equation provides a resummed function $D^R$, which in principle should work where the expansion in $b_T$-space can be treated perturbatively~\cite{Echevarria:2012pw}.
When  the values of $b_T$-coordinate increase  the convergence of $D^R$ is spoiled. This causes problems because, in order to Fourier anti-transform to the momentum space, one should know this function on the whole coordinate space, and so also in $b_T$-intervals where the function is not perturbatively calculable.
The only available strategy is therefore to check whether, within the data set at our disposal, the information that we are missing on $D^R$ is relevant or not.

The  complete perturbative limit includes the operator product expansion of the TMDPDF onto a PDF, via a matching Wilson coefficient. In other words, for small $ \vecb{b}_T$ we can write
\begin{align}
\label{eq:f1}
\tilde F_{q/N}^{\rm pert}(x,b_T;\zeta,\mu)\sim\left(\frac{\zeta}{\zeta_b}\right)^{-D(b_T;\mu)}\sum_j\int_x^1
\frac{d \tau}{\tau}\tilde C^{\not Q}_{q\leftarrow j}(x/\tau,b_T;\zeta_b,\mu)f_{j/N}(\tau,\mu)\,,
\end{align}
where $f_{j/N}$ are standard PDFs.

The choice of scales is crucial in order to properly resum the perturbative logarithms and parameterize the non-perturbative part of the TMDs correctly. The $\mu$ scale is expected to minimize the logs in the evolution kernel and it also appears in the definition of the PDF. A sensible choice is $\mu=Q_0+q_T$: at high $q_T$ the logs in the evolution kernel are certainly minimized, while a value of $Q_0\sim 2$ GeV allows to use the PDFs extracted in the literature in a natural way.
The drawback of this choice is that at high values of $b_T$ the evolution kernel becomes perturbatively unstable, although the whole evolution factor is numerically very small. It is only by confronting with data that one can establish whether this instability needs to be cured or, on the contrary, the overall suppression makes it an irrelevant problem. It is also possible that the perturbative series, although convergent, does not provide a correct value for the evolution kernel: this once again should be clarified via a comparison with experiment.
Another advantage of this scale choice is that one can avoid hitting the Landau pole in the strong coupling. Using this choice we can write
\begin{align}
\label{eq:fpert}
{\tilde F}_{q/N}^{\rm pert}(x,b_T;\z,\m) &=
\exp\le\{
\int_{\m_0}^{\m} \frac{d\bar\m}{\bar\m}
\g_V\le(\as(\bar\m),\ln\frac{\z}{\bar\m^2} \ri)\ri\}\,
\le(\frac{\z}{\zeta_b^2}\ri)^{-D^R(b_T;\m_0)}
\nn\\
&\times
\sum_{j=q,\bar q, g}
{\tilde C}_{q/j}(x_A,b_T;\z_b,\m_0)\otimes
f_{j/N}(x_N;\m_0)\,
\,.
\end{align}
Defining $\mu_b$  as $\mu_b^{-1}= b_T \ e^{\gamma_E}/2$, where $\gamma_E$ is the Euler Gamma constant, we choose
 $\z_b=C_\z^2 \m_b^2$ and $\m_0=Q_0+ q_T$.
Notice that we have explicitly kept the dependence on the real parameter $C_\z$, which will be used later on to test the dependence of the results on the rapidity scale, basically varying it between $1/2$ and $2$.
The Wilson coefficients  ${\tilde C}_{i/j}$ also obey an RG equation
\begin{align}
\frac{d}{d\ln \mu} \tilde{C}_{q/j}(x,b_T;C_\z^2\m_b^2,\m) &=
(\G_{\rm cusp} L_T - \g^V - \G_{\rm cusp}\ln C_\z^2)
\tilde{C}_{q/j}(x,b_T;C_\z^2\m_b^2,\m)
\nn\\
&
-\sum_i\int_x^1\frac{dz}{z}
\tilde{C}_{q/i}(z,b_T;C_\z^2\m_b^2,\m)\, {\cal P}_{i/j}(x/z)
\,,
\end{align}
where $L_T=\ln(\m^2/\m_b^2)$ and ${\cal P}_{i/j}(x/z)$ are the usual DGLAP splitting kernels. The double logarithms can be partially exponentiated~\cite{D'Alesio:2014vja}:
\begin{align}\label{eq:exponentiation}
\tilde{C}_{q/j}(x,b_T;C_\z^2\m_b^2,\m) \equiv
\exp\le[h_\G(b_T;\m)-h_\g(b_T;\m, C_\zeta)\ri]
{\tilde I}_{q/j}(x,b_T;\m)
\,,
\end{align}
where
\begin{align} \label{eq:h}
\frac{d h_\G}{d\ln \m} &=
\G_{\rm cusp} L_T
\,, \quad\quad
\frac{d h_\g}{d\ln \m} =
\g^V + \G_{\rm cusp} \ln C_\z^2
\,.
\end{align}
Choosing $h_{\G(\g)}(b_T;\m_b)=0$, the first few coefficients for the perturbative expansions of $h_{\G(\g)}$ are:
\begin{align}\label{hcoeffs}
h_{\G(\g)}&=\sum_n h_{\G(\g)}^{(n)} \left( \frac{\a_s}{4\pi}\right)^n
\,,
\nn\\
h_\G^{(1)} &=
\frac{1}{4} L_T^2 \G_0\,,\quad\quad\quad\quad
h_\G^{(2)} =
\frac{1}{12} (L_T^3 \G_0\b_0+3L_T^2 \G_1)\,,\nn\\
h_\G^{(3)}&=
\frac{1}{24}(L_T^4 \G_0\b_0^2+2L_T^3 \G_0\b_1
+4L_T^3 \G_1\b_0+6 L_T^2 \G_2)\,,
\nn\\
h_\g^{(1)} &=
\frac{\g_0+\G_0 \ln C_\z^2}{2 \b_0} \le(\b_0 L_T \ri)
\,,
\nn \\
h_\g^{(2)}& =
\frac{\g_0+\G_0 \ln C_\z^2}{4\b_0} \le(\b_0 L_T \ri)^2
+ \le(\frac{\g_1+\G_1 \ln C_\z^2}{2\b_0} \ri) \le(\b_0 L_T \ri)\,,\nn\\
h_\g^{(3)} &=
\frac{\g_0+\G_0 \ln C_\z^2}{6\b_0} \le(\b_0 L_T \ri)^3
+ \frac{1}{2}\le(\frac{\g_1+\G_1 \ln C_\z^2}{\b_0}
+ \frac{1}{2}\frac{(\g_0+\G_0 \ln C_\z^2)\b_1}{\b_0^2}  \ri)
\le(\b_0 L_T \ri)^2
\nn \\ & +
\frac{1}{2} \le(\frac{\g_2+\G_2 \ln C_\z^2}{\b_0}\ri) \le(\b_0 L_T \ri)
\,.
\end{align}
Resumming  this series as in~\cite{D'Alesio:2014vja} we  have
\begin{align}\label{eq:hg}
h_\g^R(b_T;\m,C_\zeta) &=
-\frac{\g_0+\G_0\ln C_\z^2}{2\beta_0}\ln(1-X)\nn \\ &
+ \frac{1}{2}\le(\frac{a_s}{1-X}\ri) \le[
- \frac{\beta_1(\g_0+\G_0\ln C_\z^2)}{\beta_0^2} (X+\ln(1-X))
+\frac{\g_1+\G_1\ln C_\z^2}{\beta_0} X\ri]
\nn\\
&+ \frac{1}{2}
\le(\frac{a_s}{1-X}\ri)^2\le[
\frac{\g_2+\G_2\ln C_\z^2}{2\beta_0}(X (2-X))\ri.
\nn \\
& \le.
+\frac{\beta_1(\g_1+\G_1\ln C_\z^2)}{2 \beta_0^2} \le( X (X-2)-2 \ln (1-X)\ri) \ri.
\nn\\
&\le.
+\frac{\beta_2(\g_0+\G_0\ln C_\z^2)}{2\beta_0^2} X^2
+\frac{\beta_1^2(\g_0+\G_0\ln C_\z^2)}{2\beta_0^3} (\ln^2(1-X)-X^2)\right]
\,,
\end{align}
where again $a_s=\as/(4\pi)$ and $X=a_s\b_0L_T$.
This scale appears both in the matching of the TMDPDF with the PDF and in the evolution ratio $\zeta/\zeta_b$ in Eq.~(\ref{eq:f1}), and it is in principle independent of the evolution kernel scale,  see e.g.~\cite{Neill:2015roa,deFlorian:2011xf}.
Summing up, for the implementation of the TMD formalism in the differential cross section these three scales represent the main source of perturbative error (see Eq.~(\ref{eq:fpert})): the $\zeta\sim Q^2$ scale is the hard scale, $\mu$ is the renormalization and/or  evolution scale, and $\zeta_b$ can be identified as the rapidity scale in~\cite{Neill:2015roa} or the resummation scale in~\cite{deFlorian:2011xf}. We remark here that a breaking of the convergence in the $\mu_b$ perturbative series would just mean that the TMDPDF cannot be split into a coefficient and a PDF for the data set under study, but would not imply the breaking of the whole TMD formalism.

For small $q_T$,  or  at high $b_T$, we expect to include corrections to the asymptotic limit so far discussed. In Ref.~\cite{D'Alesio:2014vja} we have implemented them as
\begin{align}
\tilde F_{q/N}(x,b_T;\zeta,\mu)=\tilde F_{q/N}^{\rm pert}(x,b_T;\zeta,\mu)\tilde F_{q/N}^{\rm NP}(x,b_T;\zeta)\,,
\label{eq:f2}
\end{align}
with some non-perturbative function $ F_{q/N}^{\rm NP}(x,b_T;\zeta)$.
For the non-perturbative part of the TMDPDF we used
\begin{align}
\label{eq:fgfg}
{\tilde F}^{\rm NP}_{q/N}(x,b_T;Q_i)\equiv
{\tilde F}^{\rm NP}_{q/N}(x,b_T)\left(\frac{Q_i^2}{Q_0^2}\right)^{-D^{\rm NP}(b_T)}
\,.
\end{align}
so that the non-perturbative contribution is parameterized  in the same way as the evolution kernel in Eq.~\eqref{eq:tmdkernel}.
We have studied several parameterizations of the non-perturbative part (Gaussian, polynomial, etc.) and the final one which better provides a good fit of the data, with the minimum set of parameters and $D^{\rm NP}=0$, is
\begin{align}
\label{eq:FqN3}
{\tilde F}^{\rm NP}_{q/N}(x,b_T;Q) &=
e^{-\l_1 b_T}\le(1+\l_2 b_T^2\ri)
\,.
\end{align}
It is important to emphasize that the data for $Z$-boson production are basically sensitive just to the parameter $\l_1$, that is to the exponential factor and not to the power-like term that, controlling the large-$b_T$ region, is more sensitive to small-$q_T$ data.
The two non-perturbative parameters cannot be fixed by using $Z$-boson production data alone. It is only including the low-energy DY experimental results that we can achieve a determination of both parameters. Notice that in this model we have not considered  non-perturbative corrections to the evolution kernel. This is because actual data are poorly sensitive to it.
A more complete discussion on this point can be found in~\cite{D'Alesio:2014vja}. As a result the evolution kernel used in this work is model independent.

In table \ref{tab:resummation} we collect the perturbative orders of each piece entering  the logarithmic resummation.

\begin{table}[h]
\begin{center}
\begin{tabular}{|c|c|c|c|c|c|c|c|}\hline\hline
Order & $H$ & $\hat C_{q\leftarrow j}$ & $\G_{\rm cusp}$ & $\g^V$ &
$D^R$ & $h_\G^R$ & $h_\g^R$
\\
\hline
LL &$\as^0$&$\as^0$&$\as^1$&$\as^0$&$\as^0$&$\as^{-1}$&$0$
\\
\hline
NLL &$\as^0$&$\as^0$& $\as^2$ & $\as^1$&$\as^1$&$\as^0$&$\as^0$
\\
\hline
NNLL&$\as^1$&$\as^1$&$\as^3$&$\as^2$&$\as^2$&$\as^1$&$\as^1$\\
\hline\hline
\end{tabular}
\caption{Perturbative orders in logarithmic resummations.\label{tab:resummation}}
\end{center}
\end{table}

\section{ Error analysis for $Z$-boson production}
\label{sec:Z}

\begin{figure}[t]
 \includegraphics[width=.49\textwidth, angle=0,natwidth=610,natheight=642]{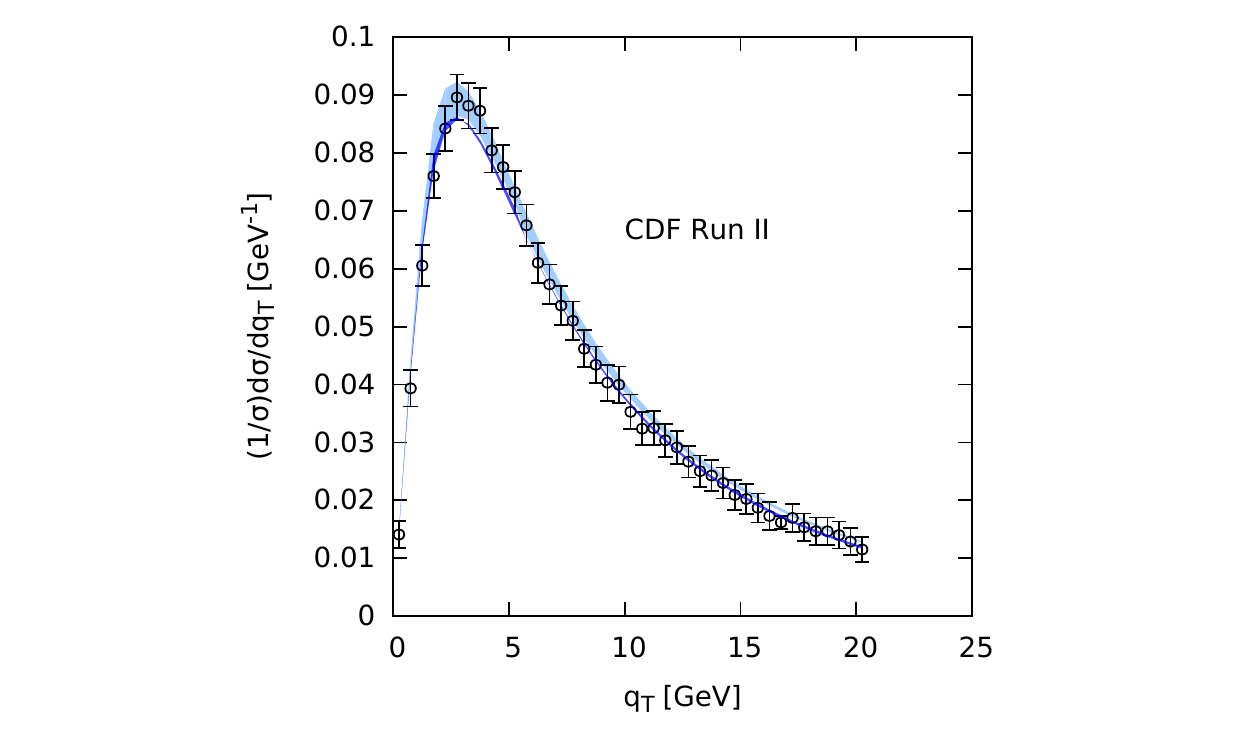}
\includegraphics[width=.49\textwidth, angle=0,natwidth=610,natheight=642]{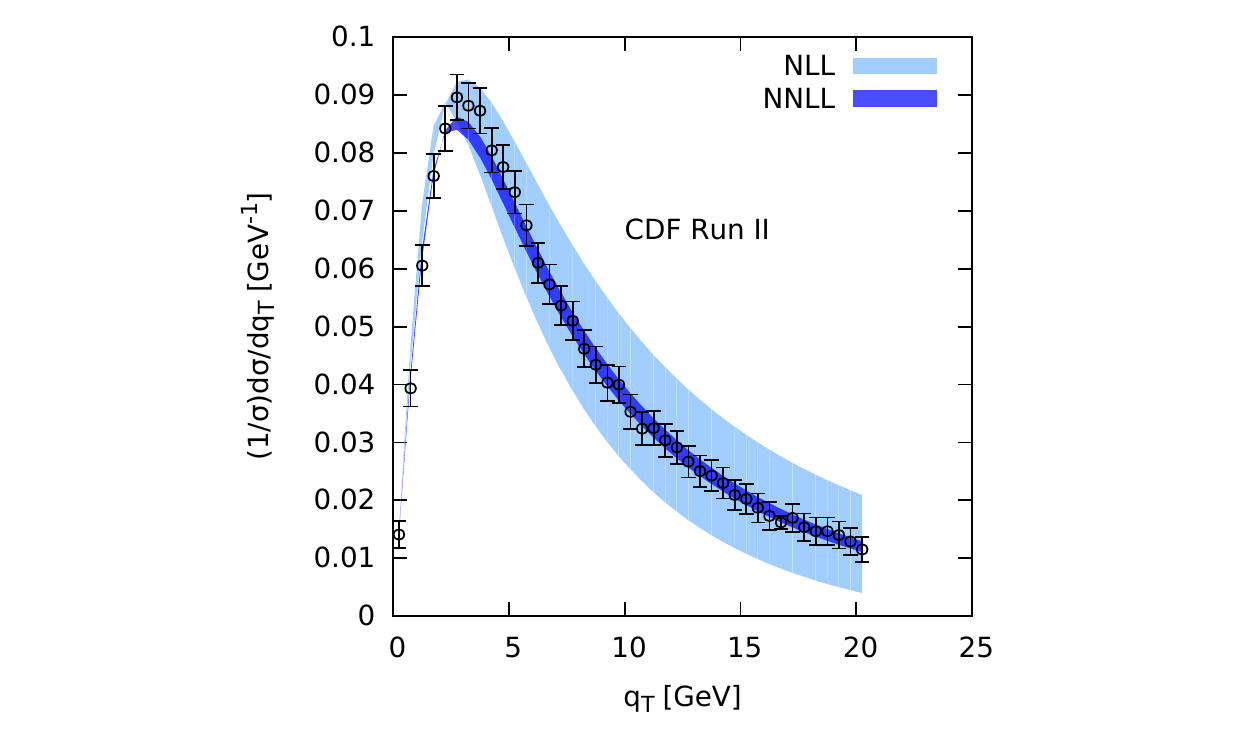}
 \caption{%
CDF Run II data for $Z$-boson production with $\mu$  (left panel) and $\mu_b$  (right panel) error  bands at NLL and NNLL. The non-perturbative part of the TMDPDF  is included, as  in Eq.~(\ref{eq:FqN3}). The values of the non-perturbative parameters are given in Ref.~\cite{D'Alesio:2014vja}.
\label{fig:CDFRII}}
 \end{figure}

\begin{figure}[ht]
 \begin{center}
\includegraphics[width=.49\textwidth, angle=0,natwidth=610,natheight=642]{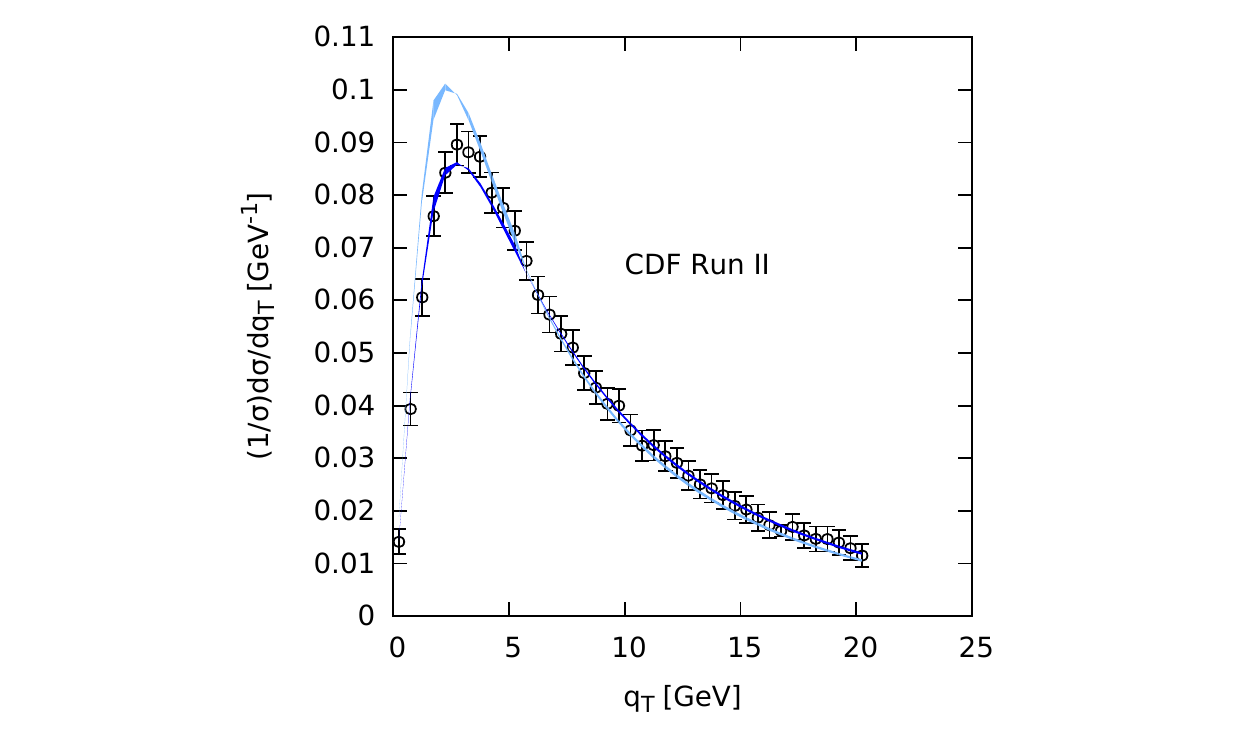}
\includegraphics[width=.49\textwidth, angle=0,natwidth=610,natheight=642]{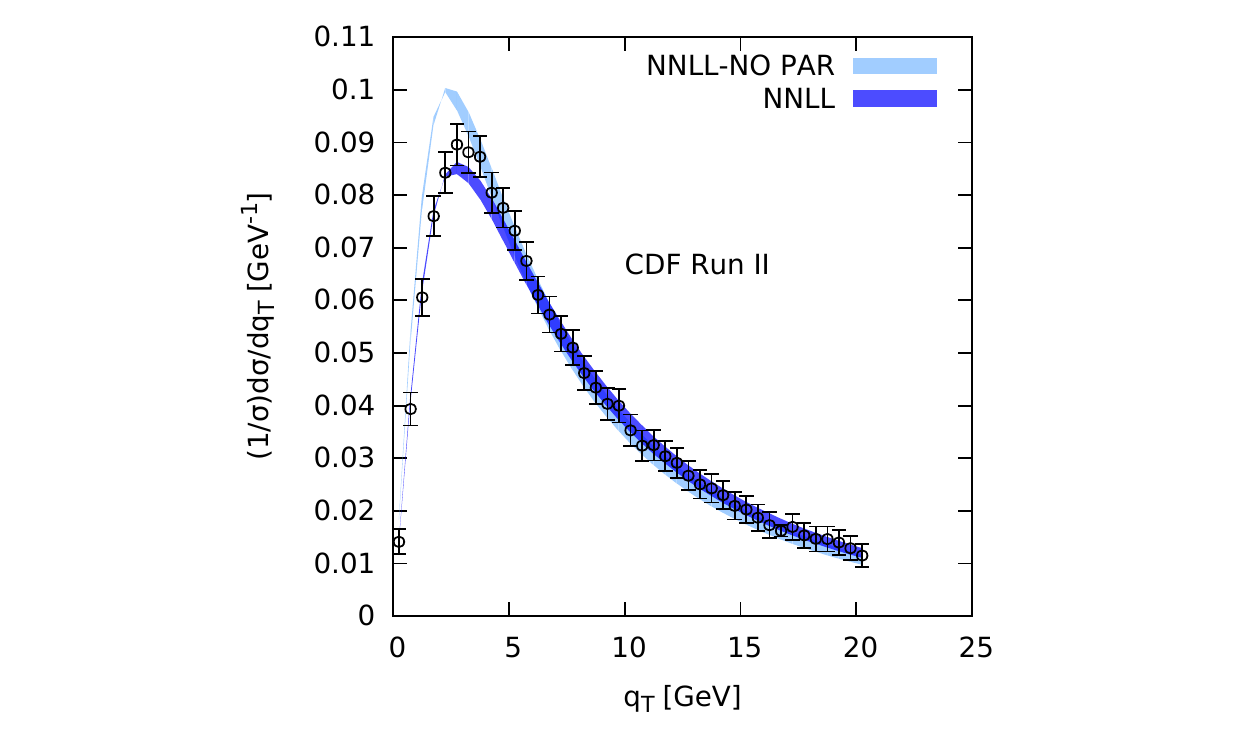}
\caption{CDF Run II data for $Z$-boson production with $\mu$  (left panel) and $\mu_b$  (right  panel) error  bands at NNLL with and without the non-perturbative part of the TMDPDF, Eq.~(\ref{eq:FqN3}). The values of the non-perturbative parameters are given in Ref.~\cite{D'Alesio:2014vja}.
\label{fig:CDFRIIv}}
 \end{center}
 \end{figure}

The error induced on $Z$-boson production data from the variation of the hard scale, $\zeta$, is found negligible and we do not discuss it in the following.
The important theoretical errors come from the $\mu$ and $\zeta_b$ variations by a factor 2 around their corresponding central values.
In Figs.~(\ref{fig:CDFRII}) and~(\ref{fig:CDFRIIv}) we show the respective bands at NLL and NNLL for CDF-Run II data.
Improving the perturbative order  has a great impact on the precision of the final result, see Figs.~(\ref{fig:CDFRII}). In both plots in Figs.~(\ref{fig:CDFRII})  the  theoretical  value  includes a non-perturbative  scale independent input, as given in Eq.~(\ref{eq:FqN3}). The values of the  parameters are the ones obtained in the fit of Ref.~\cite{D'Alesio:2014vja}. By comparing the NLL and NNLL bands it is evident that the perturbative series gets stabilized only starting at NNLL.
We can also check the theoretical impact of the non-perturbative model by looking at
Figs.~(\ref{fig:CDFRIIv}). Clearly the peak region for $q_T< 5$ GeV  needs some non-perturbative contribution to be correctly described. The $Z$-boson data per se cannot fix the parameters of the non-perturbative model, because, despite all, the sensitivity to the non-perturbative information is not so pronounced in the corresponding regime. Basically a one-parameter  model can reproduce these data~\cite{Becher:2011xn,D'Alesio:2014vja}.  In our description the non-perturbative model is actually fixed  by low energy Drell-Yan data.  Moreover   before drawing any definite conclusion about the need of non-perturbative physics to describe the peak region one has to perform a complete analysis at NNLL'/NNLO~\cite{D'Alesio:2016}.

\section{ Error analysis for DY data}
\label{sec:DY}

\begin{figure}[ht]
 \begin{center}
 \includegraphics[width=.58\textwidth, angle=0
]{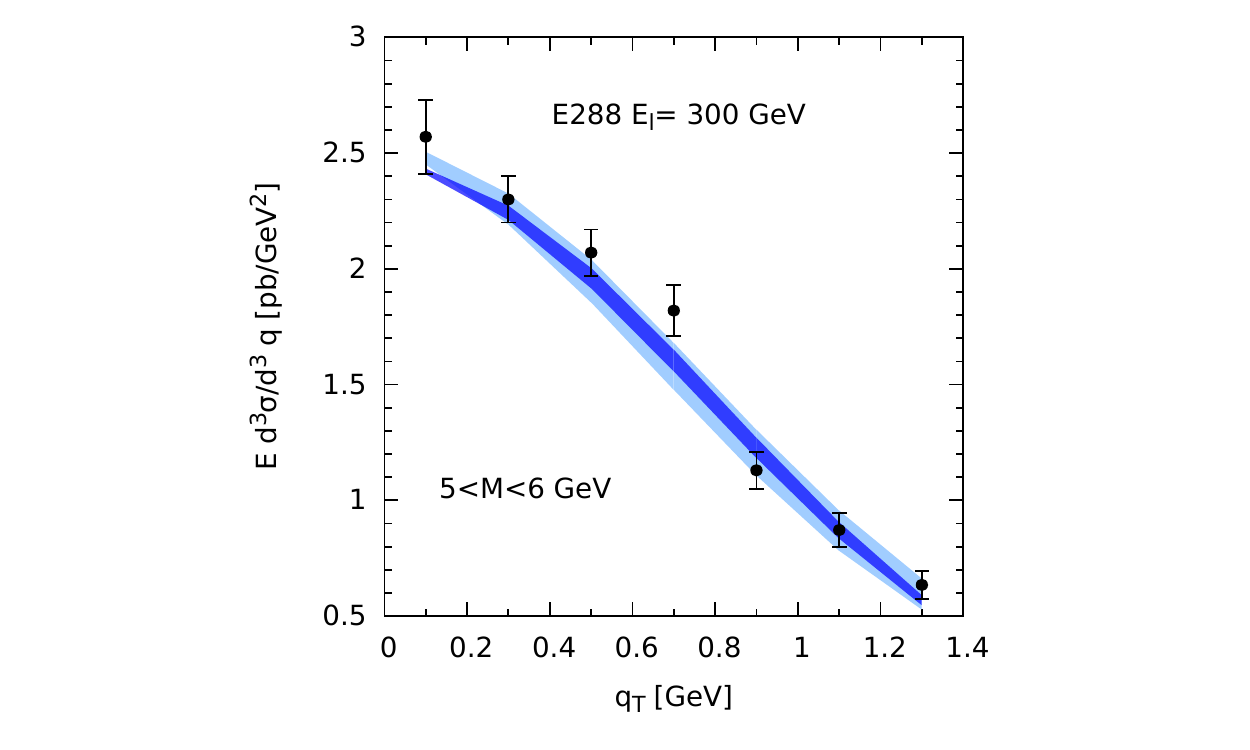}\hspace{-3cm}
\includegraphics[width=.58\textwidth, angle=0
]{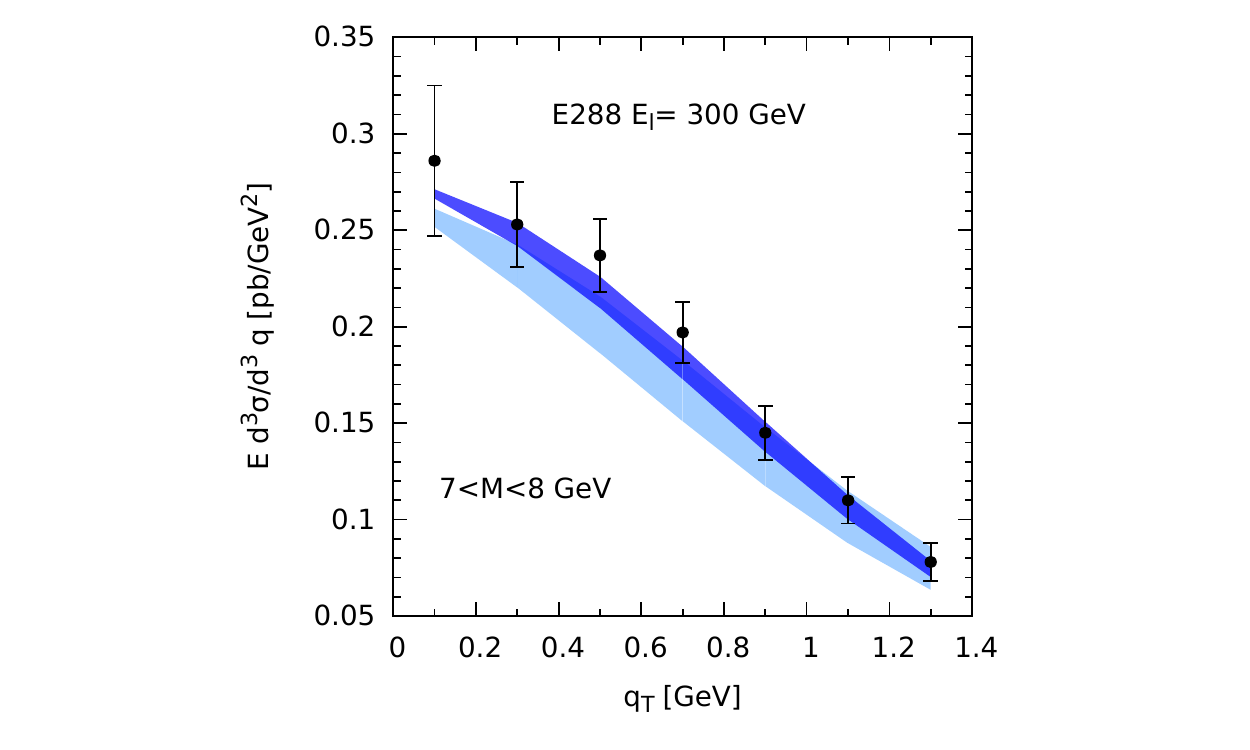}
\includegraphics[width=.58\textwidth, angle=0
]{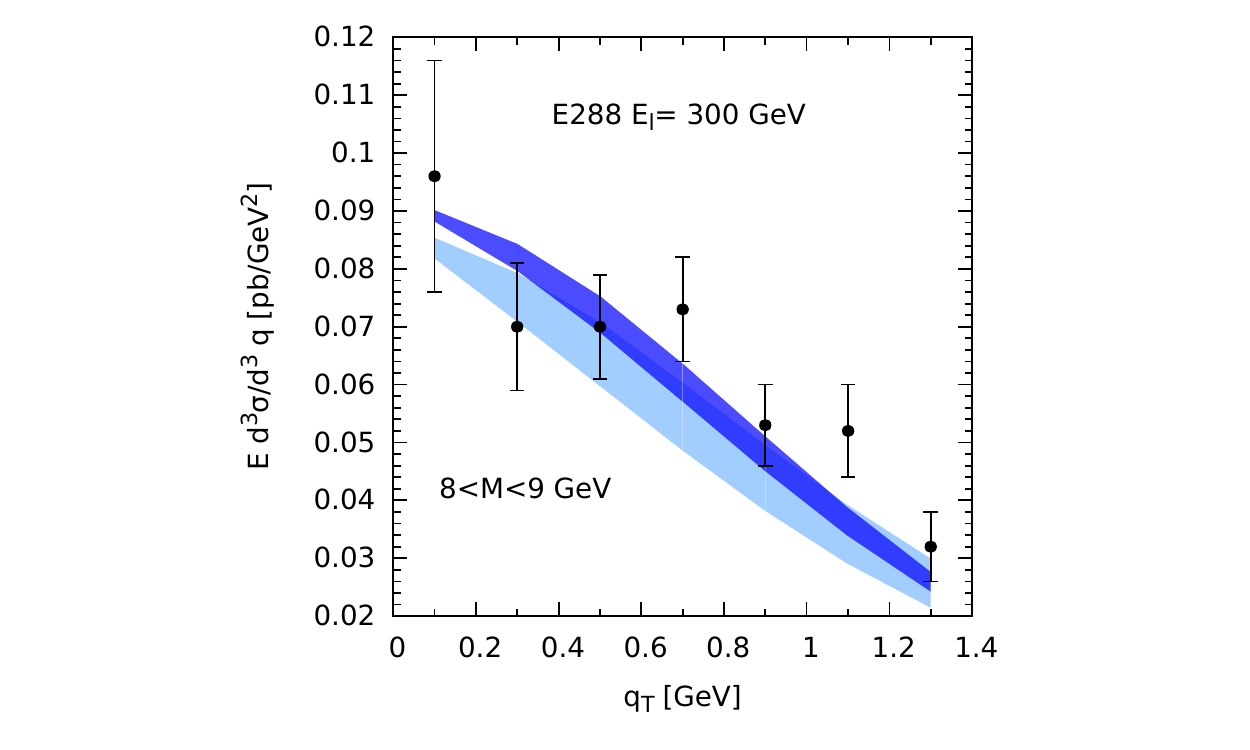}
 \caption{E288 data with $\mu$ error bands at NLL and NNLL.
\label{fig:E288}}
 \end{center}
 \end{figure}

The available data on DY have an energy generally much lower than the $Z$-boson mass and cover dilepton invariant masses above 4 GeV.
They represent a good set of data to test the TMD formalism. A non-perturbative input is now necessary and the model proposed in Eq.~(\ref{eq:FqN3}) results to give a very good description of data with a $\chi^2/d.o.f.\sim 1$. A complete study of all DY experimental results and their relative $\chi^2$  can be found in Ref.~\cite{D'Alesio:2014vja}.
The relevant point  here is the theorerical error induced  by the perturbative analysis.
The error from the hard scale, ($\zeta$), is found again negligible and we do not discuss it further.
The evolution scale error or $\mu$ error is shown in  Figs.~(\ref{fig:E288}) for one particular experiment, and similar figures can be traced for all experiments.
In general we notice a good convergence of the theory and a reduction of the error increasing the perturbative accuracy. The analysis of these two errors confirms the validity of the TMD
treatment for these data.
A perturbative analysis should be stable also when the splitting of the TMDPDF in Wilson coefficient and PDF is performed.
The rapidity ($\zeta_b$) error however is not completely under control within the perturbative expansion. While the central value  obtained with $C_\zeta=1$, coming from the perturbative series together with the non-perturbative input, provides a curve which describes well the data, a variation of the $\zeta_b$ scale in general leads to a band larger then the experimental errors.
It is then clear that the non-perturbative TMD effects here are important, although they  do not affect the  TMD evolution settings.
A higher perturbative accuracy should be implemented in order to check the limits of the perturbative expansion.

\section{Conclusions}
\label{sec:conclusions}

The transverse momentum dependent cross sections  for DY and vector boson production are known to be affected  by non-perturbative QCD effects  at small transverse momentum and/or low dilepton invariant mass. The factorization theorems for these cross sections allow to identify the elements of the cross section which should maximally include the non-perturbative effects, the TMDPDF. We have performed a study at NNLL/NNLO of the impact of the  perturbative scale variations on the fits of  DY and $Z$-boson production. The fits show that the TMD formalism can well describe the whole set of data when a NNLL/NNLO analysis is performed. In the TMD formalism the theoretical errors are mainly driven  by the  evolution scale ($\mu$) variation and marginally by the hard scale ($\zeta$) variation. 
For high values of the hard scale and transverse momentum it is possible to match the TMDs  on the integrated PDFs  and  estimate the theoretical error coming from the  rapidity scale ($\zeta_b$) variation.
We find that the decription of the peak of the $Z$-boson production needs a non-perturbative input, which  can be fixed only by fitting the low energy DY data. Increasing the perturbative order may improve the theoretical precision in this respect.

\Acknowledgements
M.G.E.~is supported by the ``Stichting voor Fundamenteel Onderzoek der Materie'' (FOM), which is financially supported by the ``Nederlandse Organisatie voor Wetenschappelijk Onderzoek'' (NWO).
U.D.~and S.M.~acknowledge support from the European Community under the FP7 program ``Capacities - Research Infrastructures'' (HadronPhysics3, Grant Agreement 283286). S.M.~is partly supported by the ``Progetto di Ricerca Ateneo/CSP'' (codice TO-Call3-2012-0103).
I.S.~is supported by the Spanish MECD grant, FPA2011-27853-CO2-02 and  FPA2014-53375-C2-2-P.

\end{document}